\documentclass[aps,prb,groupedaddress,twocolumn]{revtex4-1}

\usepackage{graphicx}

\usepackage{rotating}
\usepackage{color}
\usepackage{amssymb}

\begin{document}


\title{Ab-initio studies on the phonons of BaTiO$_{3}$ polytypes:\\ pressure dependence with a hybrid functional}


\author{Yu-Seong Seo}
\affiliation{Department of Physics, Pusan National University, Busan 609-735, Republic of Korea}

\author{Jai Seok Ahn}
\email[]{Author to whom all correspondence should be addressed.\\ Electronic address: jaisahn@pusan.ac.kr}
\affiliation{Department of Physics, Pusan National University, Busan 609-735, Republic of Korea}


\date{\today}

\begin{abstract}
We report the first principles investigations on the phonons of three polytypes of BaTiO$_3$ (BTO): paraelectric (PE) cubic $Pm$\={3}$m$ and two ferroelectric (FE) phases, tetragonal $P$4$mm$ and rhombohedral $R$3$m$. 
The phonon frequencies calculated using various exchange-correlation functionals, including density functional theory, Hartree-Fock approximation, and their hybrids were reviewed. 
The pressure-induced interplays between the modes form individual phases were explored by calculating the phonon modes as a function of pressure, $P$ from -15 to 230 GPa. 
The pressure-sensitive modes of the FE phases showed softening and converged to the modes of the PE phase at pressures below $\sim$ 10 GPa. 
These results on the FE phases can be interpreted as phonon-precursors for a change in symmetry from
low- to high-symmetry and partly as a theoretical explanation for the pressure-induced mode-coupling behaviors reported by Sood {\it et al}.~[Phys.~Rev.~B \textbf{51}, 8892 (1995)]. 
As pressure is applied further beyond $\sim$ 50 GPa to the cubic PE phase, the lowest $F_{1u}$ mode softens again and diverges into two separate modes of tetragonal FE $P$4$mm$ at above $\sim$ 150 GPa. 
These phonon-branching behaviors at high pressures provide a clear re-confirmation of the re-entrant
ferroelectricity predicted in [Phys.~Rev.~Lett. \textbf{95}, 196804 (2005); Phys.~Rev.~B \textbf{74}, 180101 (2006); \textit{ibid}. \textbf{85}, 054108 (2012)]. 
The high-pressure-re-entrant FE polarization was not found in the rhombohedral structure. 
Instead, the centosymmetric $R$\={3}$m$ phase was favored at above $\sim$ 30 GPa. 
The phonon modes calculated for the phonon-propagation vectors in the high-symmetry directions show that the $Pm$\={3}$m$ phase exhibits polar instability at the $\it\Gamma$ point and non-polar instability at the $X$, $M$, and $R$ points under high pressure.
\end{abstract}


\maketitle

\section{Introduction}

Ferroelectric (FE) materials are used for a wide range of modern technologies targeted for storage and motion, such as in memory devices, electromotive transducers, and sensors for medical imaging.\cite{Hill1988,Haertling1999} 
The two essential features of FE materials, electrostatic polarization for storage and piezoelectricity for motion, are closely related to the structures belonging to subgroups of the non-centosymmetric piezoelectric group.\cite{Haertling1999} 
FE perovskite oxides, such as Pb(Zr,Ti)O$_3$ (PZT), have three phases, rhombohedral, tetragonal, and cubic. Ferroelectricity occurs at two non-centosymmetric phases, tetragonal and rhombohedral, which are
connected to a morphotropic phase boundary (MPB) composed of a monoclinic/orthorhombic phase.\cite{Shirane1952,Jaffe1954,Jaffe1971,Noheda2000,Damjanovic2010} 
The existence of an intermediate phase is understood to be indispensible for the large piezoelectricity and polarization rotation mechanism across the MPB, as explained by Fu and Cohen,\cite{Fu2000} such as in PZT or Ba(Ti,Zr)O$_3$-\textit{x}(Ba,Ca)TiO$_3$, (BTZ-\textit{x}BCT).\cite{Damjanovic2010,Liu2009} 
More recently, evidence for local structures in relaxor-ferroelectrics, such as BTZ-\textit{x}BCT,\cite{Jeong2010,Jeong2012,Seo2013} has been reported. 
Neutron pair-distribution-function analysis revealed notable local structural features departing from the given average crystallographic structures, such as local Ti distortions toward $\langle$111$\rangle$, trigonal 3:3 Ti-O length distribution, and single bond-length of Zr-O in BTZ-\textit{x}BCT.\cite{Jeong2010,Jeong2012} Raman spectroscopy of BTZ/BTZ-\textit{x}BCT also showed rich interplay between long-range average structure and short-range local orders as a function of pressure.\cite{Seo2013,Kreisel2004} 
The phonon-interference effect in the pressure-dependent Raman spectra of BaTiO$_3$ (BTO) was explained by the coupled-phonon interaction of three $A_1$ modes of the tetragonal phase.\cite{Sood1995} 
The off-center displacements of Ti ions in the cubic phase of BTO were determined from nuclear magnetic resonance.\cite{Zalar2004}

FE BTO has been investigated many times using a first-principles method.\cite{King-smith1994,Ghosez1999,Bilc2008,Wang2010,Evarestov2012,Zhong1995,Zhang2006,Wu2005,Kornev2005,Bousquet2006,Duan2012} 
King-Smith and Vanderbilt correctly predicted the symmetry of ground-state structure as rhombohedral.\cite{King-smith1994} 
In Zhong \textit{et al}.'s pressure-temperature phase diagram, the sequence of the low temperature phases with increasing temperature was rhombohedral-orthorhombic-tetragonal-cubic.\cite{Zhong1995} 
Wu \textit{et al}.~obtained the phonon frequencies of a rhombohedral phase under a ``negative fictitious
pressure'' to constrain the cell volume to be that of the experiment.\cite{Wu2005} 
The pressure-dependence of BTO or related FE oxides, however, has rarely been studied using a first-principles method until now, except for a few recent reports. Kornev \textit{et al}.~and Bousquet 
\textit{et al}.~analyzed the re-entrant ferroelectric instability of PbTiO$_3$ \cite{Kornev2005} and BTO \cite{Bousquet2006} at high pressure, respectively. 
Duan \textit{et al}.~examined the pressure effects on the ferroelectricity and piezoelectricity of tetragonal BTO.\cite{Duan2012} 
In addition, recent successful applications of hybrid-functionals in the \textit{ab}-\textit{initio} phonon
calculations of various materials\cite{Bilc2008,Wahl2008,Goffinet2009,Hummer2009,Evarestov2011,Hong2012}
prompted us to a reexamination of the dynamic properties of BTO polymorphs as a function of pressure using a hybrid functional.
In this paper, we describe the structures and phonons of BTO polymorphs using a first-principles method in two directions. 
The first part compares the phonons of three polymorphs of BTO, paraelectric cubic and two FE phases, tetragonal and rhombohedral, calculated by using various exchange-correlation functionals including hybrid methods. 
The second part discusses the pressure dependences of the phonon modes calculated for phonon-propagation vectors in the high-symmetry directions of the three BTO polytypes.

\section{Computational Details}
First-principles calculations were performed using the CRYSTAL09 code.\cite{Dovesi2005} 
The basis sets (BSs) composed of Gaussian type orbitals were used. 
To describe heavy elements, such as barium and titanium, the Hay-Wadt effective core potentials with
small-cores were adopted for the BSs. 
The Ba (5$s$/5$p$/6$s$) and Ti (3$s$/3$p$/4$s$/3$d$) electrons were considered to be valence electrons for self-consistent calculations.\cite{Piskunov2004}  Therefore, the computational time could be minimized within our computational resources. 
For oxygen atoms, the full electron BS, O-411d11G, was used.\cite{Valenzano2006} 
The reciprocal space integration was approximated by sampling the Brillouin zone with a 6$\times$6$\times$6 mesh of the Monkhorst-Pack scheme. 
The self-consistent cycles were repeated until convergence was reached within a tolerance of 10$^{-10}$ Ha ($\sim$ 3$\times$10$^{-9}$ eV) of the total energy difference.

The ground state configurations calculated using a number of different exchange-correlation functionals, such as density functional theory (DFT), Hartree-Fock (HF) approximation, and their hybrids, were compared. 
This comparison allowed us to select the exchange-correlation functional, which best describes the lattice
structural and electronic properties of the three different structures of BTO.\cite{Seo1368} 
For DFT, a local density approximation (LDA) and generalized gradient approximations, GGA\cite{Perdew1996} and PWGGA,\cite{Perdew1992} were used. 
Four different hybrid-functional forms were applied to obtain a more accurate band gap: B3LYP,\cite{Dovesi2005} B3PW,\cite{Becke1993} PBE0,\cite{Adamo1999} and B1WC.\cite{Bilc2008} 
The B3LYP has a Becke's three parameter hybrid functional form,\cite{Becke1993} 
which includes HF exchange and the non-local correlation reported by Lee, Yang, and Parr,\cite{Lee1988}  with \textit{A} = 0.2, \textit{B} = 0.9, and \textit{C} = 0.81:
\begin{widetext}
\begin{equation}
E_{xc} = (1 - A) \cdot (E_x^{LDA} + B \cdot E_x^{Becke}) + A \cdot E_x^{HF} + (1-C) \cdot E_c^{VWN} + C \cdot E_c^{LYP},
\label{Eq1}
\end{equation}
\end{widetext}

\noindent 
where $E_x^{LDA}$ and $E_c^{VWN}$ are the LDA exchange and Vosko-Wilk-Nusair correlation, \cite{Vosko1980} respectively, which fits the Ceperley-Alder data for an electron gas.\cite{Ceperley1980}  Similarly, the B3PW hybrid functional uses PWGGA, $E_c^{PWGGA}$, as a non-local correlation instead of
$E_c^{LYP}$ in B3LYP shown in Eq.~(1). 
The PBE0 is a parameter-free hybrid functional, which uses the PBE (GGA) exchange-correlation mixed with 25\% of HF exchange. 
The B1WC is similar to the PBE0 but mixes 16\% of the HF exchange on Wu-Cohen-GGA\cite{Wu2006} exchange-correlation, which is tuned to remove the super-tetragonality within the GGA-derived hybrid
functionals, as will be discussed in Sec.~III.

The equilibrium geometries were determined by optimizing the lattice constants and fractional displacements
of the basis atoms for the cubic ($Pm$\={3}$m$), tetragonal ($P$4$mm$), and rhombohedral ($R$3$m$) structures of BTO: structural relaxations were repeated until the norm of the total forces became less than 3$\times$10$^{-5}$ Ha/Bohr ($\approx$ 1.5$\times$10$^{-3}$ eV/\AA). 
The phonon modes were calculated using a frozen phonon method implemented in CRYSTAL09 code. 
The phonon dispersions were calculated using a supercell method: a 2$\times$2$\times$2 supercell that consists of 40 atoms was assumed for each one of the cubic, tetragonal, and rhombohedral (i.e., 2$\times$2$\times$1 in hexagonal setting) structures. 
The electronic dielectric constants and Born effective charges were also calculated to obtain LO/TO splitting of the zone-center phonon by a non-analytic correction, which is due to the long range electrostatic interactions in polar materials. 
The high frequency (optical) dielectric constants were calculated using the finite field perturbation method: the effect of a ``sawtooth'' electric potential applied in a supercell was evaluated numerically. 
For the tetragonal or rhombohedral (in hexagonal setting) case, the anisotropic dielectric constants, $\varepsilon_{a,\infty}$ and $\varepsilon_{c,\infty}$, were determined by applying an electric field through the long axes of the supercells, respectively, taken along the crystallographic $a$ and $c$ directions. The Born effective charges were evaluated using the Berry phase approach.\cite{King-Smith1993}

\section{Results and Discussion}

\begin{table*}[ht]
\caption{%
Calculated lattice parameters (in \AA), dielectric constants, and Born-effective-charge tensors ($\textit{\textbf{Z}}^*$ in $|e|$) of the cubic, tetragonal, and rhombohedral phases of BaTiO$_3$.
The present calculations using the B1WC and LDA functional are compared with previous LDA results calculated at the experimental volumes. 
Experimental values are from Refs.~46-50.}
\begin{ruledtabular}
\begin{tabular}{ccccccccccccccc}
&\multicolumn{4}{c}{Cubic
\textit{Pm}\={3}\textit{m}}&&\multicolumn{4}{c}{Tetragonal
\textit{P}4\textit{mm}}&&\multicolumn{4}{c}{Rhombohedral
\textit{R}3\textit{m}}
\\\cline{2-5}\cline{7-10}\cline{12-15} & B1WC & LDA
&LDA\footnote[1]{Ref.~17.} & Expt. && B1WC & LDA
& LDA\footnote[2]{Ref.~45.} & Expt. && B1WC & LDA & LDA\footnotemark[2] & Expt. \\
*& at 0 K & at 0 K & at 300 K & at 300 K && at 0 K & at 0 K & at 300 K & at 300 K && at 0 K & at 0 K & at 300 K & at 300 K \\
\hline
\textit{a}&3.952&3.936&4.000&3.996\footnote[3]{Ref.~46.}&&\textit{a}: 3.958&3.949&3.994&3.994\footnote[5]{Ref.~48.}&&\textit{a}: 3.991&3.962&4.003&4.004\footnotemark[5] \\
&&&&&&\textit{c}: 4.042&3.984&4.036&4.036\footnotemark[5]&&$\alpha$: 89.86$^{\circ}$ &89.96$^{\circ}$&89.84$^{\circ}$&89.84$^{\circ}$\footnotemark[5] \\
&&&&&&$c/a$: 1.021&1.009&1.011&1.011\footnotemark[5]&&&&& \\
\hline
\textit{V}&61.72&60.97&64.00&63.81\footnotemark[3]&&63.34&62.15&64.38&64.38\footnotemark[5]&&63.55&62.20&64.14&64.17\footnotemark[5] \\
\hline
$\varepsilon_{\infty}$&5.12&5.81&6.75&5.40\footnote[4]{Ref.~47.}&&\textit{a}: 4.92&5.69&6.48&5.93\footnote[6]{Ref.~49.}&&4.96&6.01&6.16&6.19\footnote[7]{Ref.~50.} \\
&&&&&&\textit{c}: 4.56&5.41&5.84&5.60\footnotemark[6]&&4.60&5.95&5.73&5.88\footnotemark[7] \\
\hline
&(Ba)&&&&&&&&&&&&&\\
$\textit{\textbf{Z}}^*$&2.675&2.726&2.74&&&\textit{xx}: 2.658&2.708&2.726&&&2.711&2.742&2.783&\\
&&&&&&\textit{zz}: 2.776&2.769&2.814&&&2.681&2.696&2.737\\
\cline{2-15}
&(Ti)&&&&&&&&&&&&&\\
&7.029&7.235&7.32&&&\textit{xx}: 6.732&7.094&7.044&&&6.410&6.863&6.608&\\
&&&&&&\textit{zz}: 5.686&6.385&5.971&&&5.548&6.430&5.765\\
\cline{2-15}
&(O)&&&&&(O$_1$)&&&&&(O)&&&\\
&\textit{xx}: -5.584&-5.707&-5.76&&&\textit{xx}: -1.948&-2.061&-2.024&&&\textit{xx}: -2.489&-2.618&-2.562&\\
&\textit{zz}: -2.060&-2.127&-2.14&&&\textit{zz}: -4.596&-5.139&-4.836&&&\textit{xy}: -0.956&-1.012&-0.984&\\
&&&&&&&&&&&\textit{xz}: 0.712&0.788&0.647&\\
&&&&&&&&&&&\textit{yx}: -0.956&-1.012&-0.984&\\
&&&&&&(O$_2$)&&&&&\textit{yy}: -3.592&-3.786&-3.699&\\
&&&&&&\textit{xx}: -2.046&-2.117&-2.149&&&\textit{yz}: 1.233&1.366&1.121&\\
&&&&&&\textit{yy}: -5.396&-5.623&-5.596&&&\textit{zx}: 0.610&0.746&0.733&\\
&&&&&&\textit{zz}: -1.933&-2.008&-1.974&&&\textit{zy}: 1.057&1.293&1.269&\\
&&&&&&&&&&&\textit{zz}: -2.743&-3.042&-2.834&\\
\end{tabular}
\end{ruledtabular}
\label{table1}
\end{table*}

Table I summarize the results calculated for the relaxed geometries of the three phases of BTO by using the B1WC and LDA functionals. 
The cell volumes of the three structures calculated using the B1WC hybrid functional overestimate the LDA results, but provide better estimations of the experimental volumes. 
However, the tetragonality factor, $c/a$ = 1.021, of $P$4$mm$, which was calculated using the B1WC functional, slightly overestimates the result from LDA ($\sim$ 1.011)\cite{Hermet2009} or the experimental value ($\sim$ 1.010).\cite{Shirane1957} 
These generic behaviors regarding the lattice parameters, the successful reproduction of the cell volumes and the super-tetragonality (of $P$4$mm$), can be attributed to the non-local exchange-correlation functional embedded within the GGA-derived hybrid functionals, which is well-known for accurate predictions of the cell volumes. 
The $c/a$-ratios of the tetragonal BTO calculated using seven different exchange-correlation functionals in the previous comparative study are summarized here for references:\cite{Seo1368}  $c/a$ = 1.077 (HF), 1.009 (LDA), 1.038 (PWGGA), 1.039 (GGA), 1.072 (B3LYP), 1.046 (B3PW), and 1.042 (PBE0). 
The GGA-derived functionals (from PWGGA to PBE0), other than the Wu-Cohen-GGA, gave super-tetragonal results: $c/a$ = 1.038 - 1.072. 
These super-tetragonal results are undesirable for ferroelectric BTO. 
Therefore, the B1WC functional with $c/a$ = 1.021 was adopted for the following phonon calculations.

\begin{table*}[ht]
\caption{%
Calculated frequencies (cm$^{-1}$) of the zone-center phonon modes of the cubic, tetragonal and rhombohedral phases of BaTiO$_3$.
Present calculations using the B1WC and LDA functional are compared with previous LDA results calculated at the experimental volumes.}
\begin{ruledtabular}
\begin{tabular}{cccccccccccc}
&\multicolumn{3}{c}{Cubic
\textit{Pm}\={3}\textit{m}}&&\multicolumn{3}{c}{Tetragonal
\textit{P}4\textit{mm}}&&\multicolumn{3}{c}{Rhombohedral
\textit{R}3\textit{m}} \\\cline{2-4}\cline{6-8}\cline{10-12} &
B1WC & LDA &LDA\footnote[1]{Ref.~17.} && B1WC & LDA
& LDA\footnote[2]{Ref.~45.} && B1WC & LDA & LDA\footnotemark[2] \\
*& at 0 K & at 0 K & at 300 K && at 0 K & at 0 K & at 300 K && at 0 K & at 0 K & at 300 K \\
\hline
TO&&&&&125$i$ ($E$)&91$i$&161$i$&&125 ($E$)&144&163\\
&174$i$ ($F_{1u}$)&92$i$&219$i$&&185 ($A_1$)&186&161&&192 ($A_1$)&195&167\\
&198 ($F_{1u}$)&194&166&&190 ($E$)&188&167&&217 ($E$)&191&210\\
&312 ($F_{2u}$)&296&281&&311 ($B_1$)&302&287&&307 ($A_2$)&299&277\\
&495 ($F_{1u}$)&496&453&&313 ($E$)&297&284&&318 ($E$)&305&293\\
&&&&&334 ($A_1$)&242&302&&285 ($A_1$)&199&259\\
&&&&&488 ($E$)&487&457&&497 ($E$)&493&470\\
&&&&&543 ($A_1$)&507&507&&542 ($A_1$)&508&512\\
\hline
LO&&&&&184 ($E$)&185&162&&193 ($E$)&190&174\\
&191 ($F_{1u}$)&190&159&&203 ($A_1$)&194&180&&199 ($A_1$)&195&178\\
&480 ($F_{1u}$)&468&445&&312 ($E$)&297&284&&318 ($E$)&305&293\\
&737 ($F_{1u}$)&736&631&&473 ($E$)&463&444&&472 ($E$)&464&441\\
&&&&&485 ($A_1$)&468&452&&491 ($A_1$)&469&461\\
&&&&&724 ($E$)&720&641&&740 ($A_1$)&716&687\\
&&&&&770 ($A_1$)&735&705&&730 ($E$)&710&676\\
\end{tabular}
\end{ruledtabular}
\label{table2}
\end{table*}

The Born effective charges (BECs) were calculated using the Berry phase method for the fully relaxed geometries.
Among the calculated tensor components of the BECs using the B1WC, most are very close but slightly smaller than the LDA results for all phases. 
The electronic dielectric constants, $\varepsilon_{\infty}$'s, calculated for the three phases also underestimate the LDA predictions but match the experimental values more closely. 
The underestimations of $\varepsilon_{\infty}$'s and the $\textit{\textbf{Z}}^*$'s need to be analyzed concurrently because they affect the LO phonon modes in different ways. 
The BECs are dynamic charges that are useful for calculating the so-called LO-TO splitting through a non-analytical correction in polar materials, such as BaTiO$_3$. 
In the long-wavelength limit, the dynamical matrix, $D_{\alpha,\beta}(\textit{\textbf{k}};\mu\nu)$, of a polar crystal can be expressed as the sum of the analytical $D_{\alpha,\beta}^{(0)}(\textit{\textbf{k}};\mu\nu)$ and non-analytical contributions,\cite{Parlinski2000} such as

\begin{widetext}
\begin{equation}
D_{\alpha,\beta}(\textit{\textbf{k}};\mu\nu) = D_{\alpha,\beta}^{(0)}(\textit{\textbf{k}};\mu\nu)+\frac{4\pi e^2}{V\varepsilon_{\infty}\sqrt{M_{\mu}M_{\nu}}}
\frac{[\textit{\textbf{k}}\cdot\textit{\textbf{Z}}^*(\mu)]_{\alpha}[\textit{\textbf{k}}{\cdot}\textit{\textbf{Z}}^*(\nu)]_{\beta}}{|\textit{\textbf{k}}|^2},
\label{Eq2}
\end{equation}
\end{widetext}

\noindent
where $D_{\alpha,\beta}^{(0)}(\textit{\textbf{k}};\mu\nu)$ is the dynamical matrix derived using the direct method from the Hellmann-Feynman forces. 
In Eq.~(2), \textit{\textbf{k}} is the wave vector, $V$ is the volume of the primitive cell, and $M_{\mu}$ and $\textit{\textbf{Z}}^*(\mu)$ are atomic masses and BEC tensor of atom indexed with $\mu$, respectively. 
The non-analytic contributions are proportional to the squares of $\textit{\textbf{Z}}^*$'s but inverse-proportional to $\varepsilon_{\infty}$'s. 
Therefore, the effects from the small underestimations of the BECs are clearly overwhelmed by the (relatively) larger effect by the underestimations of $\varepsilon_{\infty}$: i.e., LO-TO splitting is dominated by the
differences in $\varepsilon_{\infty}$'s rather by the $\textit{\textbf{Z}}^*$'s.

\subsection{Optical phonon modes at ambient pressure}
In the cubic perovskite structure of BaTiO$_3$ with $Pm$\={3}$m$ symmetry, i.e. with space group (SG) No.~221, there are 12 optical modes at the $\mathit{\Gamma}$-point (i.e., at zone-center): three triply degenerate modes of $F_{1u}$ and one triply degenerate silent mode of $F_{2u}$ symmetry, with
$\mathit{\Gamma}_{optic}=3F_{1u}+F_{2u}$ as its irreducible representation at $\mathit{\Gamma}$-point.  In the tetragonal BaTiO$_3$ with $P$4$mm$ symmetry (SG No.~99), each of the $F_{1u}$ modes splits into a nondegenerate $A_1$ mode and a doubly degenerate $E$ mode, and $F_{2u}$ mode splits into $E$ and $B_1$ modes: i.e. $\mathit{\Gamma}_{optic}=3A_1+B_1+4E$, where the $E$ and $A_1$ modes are both infrared (IR) and Raman active, whereas $B_1$ is a Raman only mode. 
Similarly in rhombohedral BaTiO$_3$ with $R$3$m$ symmetry (SG No.~160), $F_{1u}$ splits into $A_1$ \& $E$ and $F_{2u}$ splits into $A_2$ \& $E$; $\mathit{\Gamma}_{optic}=3A_1+A_2+4E$, with $A_2$ being a Raman only mode.\cite{Bradley1972} 

Table II summarizes the zone-center optical phonon modes calculated using the B1WC and LDA functionals. 
The symmetries and frequencies of the phonon modes for all the three phases were obtained. 
The LO-TO splittings in the long-wavelength limit were calculated by applying the non-analytical corrections along the (100) directions. 
In agreement with previous LDA results,\cite{Ghosez1999,Hermet2009} non-stable modes with imaginary
frequencies, i.e.~soft modes signifying structural instability, were obtained for the cubic and tetragonal phases of BaTiO$_3$; the real frequencies were calculated for all the modes in the rhombohedral phase. 
The mode frequencies calculated using the B1WC and LDA functionals are comparable to the previous LDA results,\cite{Ghosez1999,Hermet2009} except for a few out-of-order TO modes at $\sim$ 300 cm$^{-1}$. A possible preliminary clue for these discrepancies, which will be proven later using the pressure-dependencies, is that the previous LDA results\cite{Hermet2009} for the tetragonal and rhombohedral structures were calculated by constraining the cell volume to the experimental value. 
These results were obtained from the fully relaxed geometry including the cell volume without any assumption.

The zone-center optical phonon modes were also calculated for the three BTO phases using various exchange-correlation functionals, such as LDA, PWGGA, PBE(GGA), B1WC, B3LYP, B3PW, PBE0, and HF. 
The results are plotted in Fig.~1. 
The closed and open symbols represent the TO modes and LO modes, respectively.
First, the results from GGA are indistinguishable from the PWGGA results. 
The three hybrid functionals provide similar results. 
The results from B3PW fully agree with the PBE0 results. 
The B3LYP provide slightly lower (in $Pm$\={3}$m$) or slightly higher (in $P$4$mm$/$R$3$m$) frequencies. 
The results from the LDA or HF deviated significantly from the results of the other functionals. 
The LDA gives out-of-ordered TO modes near 300 cm$^{-1}$ for both $P$4$mm$ and $R$3$m$. 
The HF provides no soft-modes for $P$4$mm$, whereas all the other functionals do. 
Among the GGA-derived functionals, the B1WC provides the closest values to the LDA results for both
$P$4$mm$ and $R$3$m$. 
In the $Pm$\={3}$m$ phase, shown in Fig.~1(a), the phonon modes do not vary drastically, irrespective of the functional choices.

\begin{figure}
\centering
\includegraphics[scale=0.45]{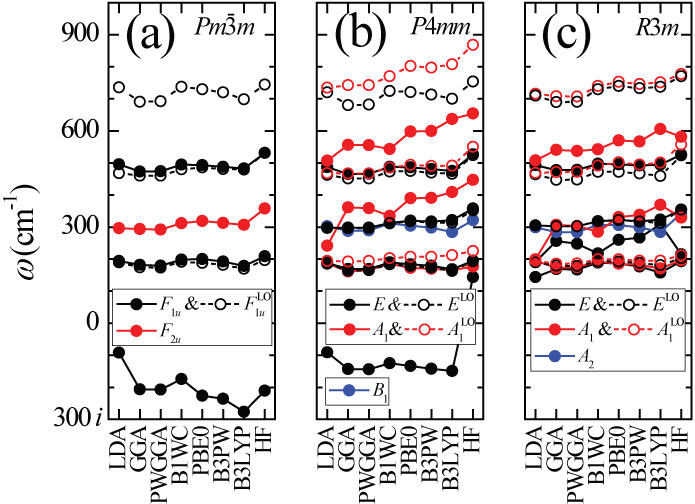}
\caption{Optical phonon modes at the $\mathit{\Gamma}$-point calculated within different functionals for (a) $Pm$\={3}$m$ cubic, (b) $P$4$mm$ tetragonal, and (c) $R$3$m$ rhombohedral phases of BaTiO$_3$. 
For $Pm$\={3}$m$: $F_{1u}$ ($\bullet$) and $F_{2u}$ ($\textcolor{red}{\bullet}$); For
$P$4$mm$: $E$ ($\bullet$), $A_1$ ($\textcolor{red}{\bullet}$), and $B_1$ ($\textcolor{blue}{\bullet}$); For $R$3$m$: $E$ ($\bullet$), $A_1$ ($\textcolor{red}{\bullet}$), and $A_2$ ($\textcolor{blue}{\bullet}$). 
Open circles designate corresponding LO modes.}
\label{fig1}
\end{figure}

\subsection{Pressure dependence of the phonon modes and structural instability}
If the phonon modes are calculated as a function of pressure, one can see the evolution and interplay of the modes. 
In addition, the connections between the irreducible representations of each structure can be understood more clearly. 
The phonon modes were calculated for the fully relaxed geometries under applied hydrostatic pressures; pressures ranged from -15 (negative fictitious pressure) to 230 GPa. 
The LDA and B1WC were selected for the pressure-dependent calculations for the three BTO phases. 
The LDA is for the comparison with other literatures. 
The B1WC hybrid functional is for the successful reproduction of the experimental cell volume, $c/a$-ratio, and energy gap.

\begin{figure}[b]
\centering
\includegraphics[scale=0.42]{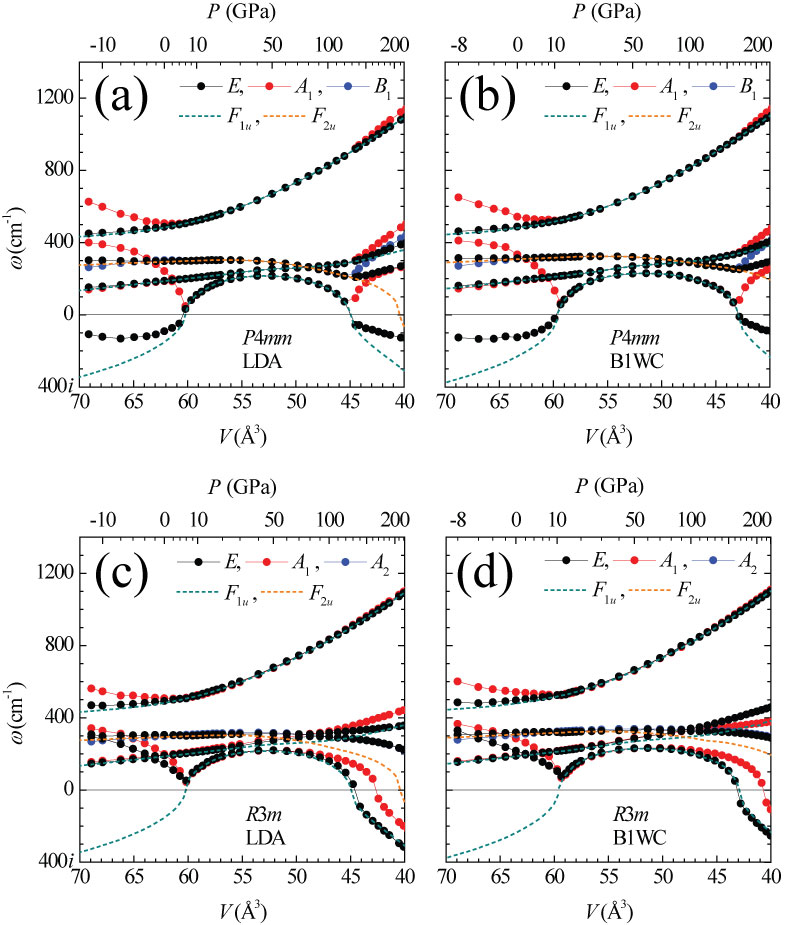}
\caption{Pressure dependences of the optical phonon modes at the $\mathit{\Gamma}$-point calculated for $P$4$mm$ with (a) LDA and (b) B1WC; for $R$3$m$ with (c) LDA and (d) B1WC. $E$ ($\bullet$), $A_1$ ($\textcolor{red}{\bullet}$), and $B_1$ ($\textcolor{blue}{\bullet}$) for $P$4$mm$; $E$ ($\bullet$), $A_1$ ($\textcolor{red}{\bullet}$), and $A_2$ ($\textcolor{blue}{\bullet}$) for $R$3$m$. 
The $F_{1u}$ and $F_{2u}$ modes calculated for the $Pm$\={3}$m$ cubic phase are overlaid with dashed lines.}
\label{fig2}
\end{figure}

The results are shown in Figure 2. 
The data points under ambient pressure (at $P$ = 0) reproduce the phonon modes tabulated
in Table II. (Note that the LO-TO splitting was not considered here.) 
Eight modes can be resolved in $P$4$mm$ (LDA: Fig.~2(a); B1WC: Fig.~2(b)) and in $R$3$m$ (LDA: Fig.~2(c); B1WC: Fig.~2(d)). 
As pressure increases from zero to $\sim$ 6 GPa in LDA or to $\sim$ 12 GPa in B1WC, most modes show  blue-shifting behaviors due to the mode Gr\"{u}neissen effect,\cite{Ashcroft1976} and the trajectories of the modes converge to one of the four modes of $Pm$\={3}$m$, three $F_{1u}$'s and one $F_{2u}$, which are shown with dashed lines on each panel. 
The most exotic behavior was found from the pressure-trajectories of the two modes, 125$i$ ($E$) and 334
($A_1$) of $P$4$mm$; 217 ($E$) and 285 ($A_1$) of $R$3$m$, tabulated for B1WC in Table II (calculated at $P$ = 0). 
In $P$4$mm$, as pressure increases, the unstable $E$ mode becomes a real-frequency mode and merges with the $A_1$ mode to become the lowest $F_{1u}$ mode at $V$ $\sim$ 60.19 \AA$^3$ (LDA) or 59.31 \AA$^3$ (B1WC). 
In $R$3$m$, however, the two stable modes ($E$ and $A_1$) showed a red-shift with decreasing volume, and merged to the lowest $F_{1u}$ lines at $V$ $\sim$ 60.16 \AA$^{3}$ for (LDA) or 59.36 \AA$^3$ (B1WC). 
The transitions to $Pm$\={3}$m$ from $P$4$mm$ or from $R$3$m$ occur at the same critical volume within a 0.1\% error using the same functional.

The calculated pressure-dependence of the phonon modes of $P$4$mm$ below $\sim$ 10 GPa can provide the long-awaited explanation based on first-principles regarding the pressure-induced phonon-coupling behaviors reported years ago by Sood \textit{et al}.\cite{Sood1995} 
According to Sood \textit{et al}., three $A_1$ (TO) modes revealed anomalous behaviors as a function of pressure for $P$ = 0-4 GPa: (i) non-monotonic pressure-dependences at $\sim$ 2 GPa, (ii) asymmetries in the line shapes, (iii) interference effects, etc. 
These behaviors were explained by the existence of a coupled-mode interaction involving only a phonon together with the structural transition from tetragonal to cubic occurring at 2.2 GPa. 
Our results, shown in Fig.~2(a) and Fig.~2(b), reproduce only some of their observations. 
The two $A_1$ modes showed (counter-intuitive) softening behavior as a function of pressure, and the pressure-rates of the modes were non-monotonic, but no peculiarity was observed at $\sim$ 2 GPa. 
Such behaviors of the two $A_1$ modes with increasing pressures, however, can be interpreted as the precursors of a symmetry change from low- to high-symmetry: the two $A_1$ modes (with two $E$ modes) merged to the $F_{1u}$ modes of the cubic phase, as indicated in the results of the cubic phase, $Pm$\={3}$m$, which is indicated with dashed lines in Fig.~2(a) and in Fig.~2(b). 
In addition, any non-linearity that may be needed for the mode-coupling model by Sood \textit{et al}.~is beyond the capability of the present calculation, because our first-principles calculation is limited to the harmonic approximation for phonon modes.

\begin{figure}[t]
\centering
\includegraphics[scale=0.4]{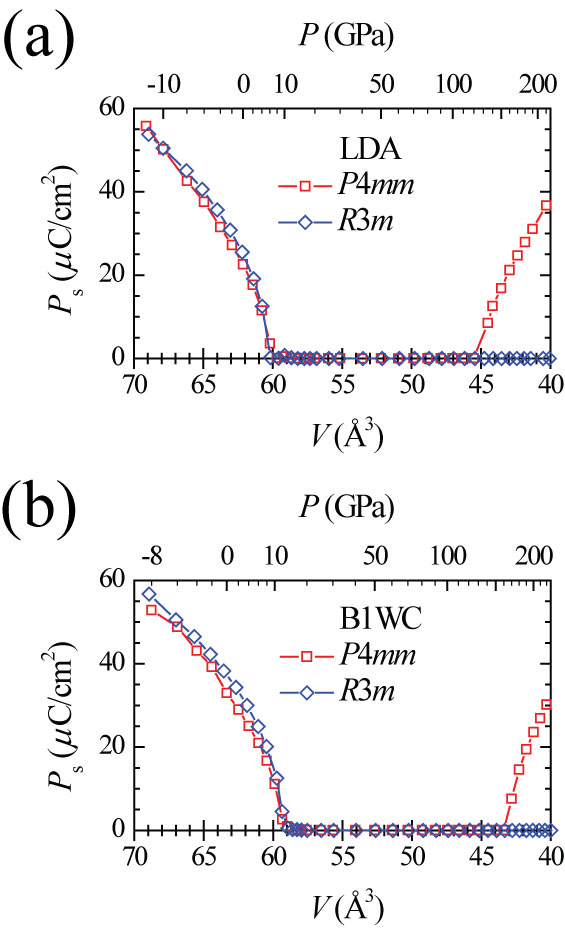}
\caption{Calculated spontaneous polarizations of the ferroelectric structures of BaTiO$_3$, $P$4$mm$ ($\textcolor{red}{\Box}$) and $R$3$m$ ($\textcolor{blue}{\diamond}$), as functions of pressure: (a) with LDA and (b) with B1WC. Polarization vectors are parallel to the [001] ($P$4$mm$) and to the [111] ($R$3$m$) directions of the pristine cubic lattice.}
\label{fig3}
\end{figure}

Let's come back to the pressure dependences of phonon modes of tetragonal structure shown in Fig.~2(a) and Fig.~2(b). 
As pressure increases beyond $\sim$ 6 GPa (LDA) or 12 GPa (B1WC), where the structure is $Pm$\={3}$m$ cubic, the lowest $F_{1u}$ mode shows hardening until it reaches a maximum at
$\sim$ 50 GPa, and then it shows a softening again until $\sim$ 120 GPa (LDA) or 160 GPa (B1WC). 
As pressure increases more above 135 GPa (LDA) or 170 GPa (B1WC), each phonon mode of the
$Pm$\={3}$m$ cubic diverges into two separate phonon modes of $P$4$mm$ tetragonal again. 
These phonon-branching behaviors provide clear re-confirmation of the re-entrant ferroelectricity under high pressure, which was first reported for PbTiO$_3$ by Kornev \textit{et al}.\cite{Kornev2005} and further
explored with the unstable cubic $F_{1u}$ mode by Bousquet and Ghosez.\cite{Bousquet2006} 
We also meticulously confirmed the re-entrant ferroelectricity by the reappearance of spontaneous
polarization and atomic displacements, which are provided in supplement materials. 
The calculated polarizations are summarized in Fig.~3(a) for LDA and in Fig.~3(b) for B1WC, respectively. 
The FE polarization disappears at $P$ $\approx$ 8 GPa (LDA) and reappears at high pressures above $P$ $\approx$ 135 GPa (LDA), which reproduces the previous results.\cite{Duan2012} 
With the B1WC functional, the FE-reappearing pressure was expected at $\approx$ 170 GPa.

In the rhombohedral structure, however, the re-entrant polarization was not found under high pressure, as shown in Fig.~3(a) and Fig.~3(b). 
The FE polarization for the rhombohedral structure under ambient pressure disappears at above $P$
$\approx$ 8 GPa (LDA), and remains at a zero level at pressures below $\sim$ 230 GPa. 
However, the cell experiences two successive symmetry-changes for the pressure range between -15 and 230 GPa: $R$3$m$ to $Pm$\={3}$m$ (cubic) at $P$ $\approx$ 8 GPa; $Pm$\={3}$m$ to $R$\={3}$m$ at $P$ $\approx$ 30 GPa. 
The rhombohedral cell at above $P$ $\approx$ 30 GPa has a centosymmetric $R$\={3}$m$ symmetry: the basis atoms are Ba (0, 0, 0), Ti (1/2, 1/2, 1/2), and O (1/2, 1/2, 0) in a rhombohedral setting; the rhombohedral angle $\alpha$ begins to deviate from 90$^{\circ}$ at pressures $\geq$ 30 GPa. 
The evolution of the structural parameters is provided separately in the supplement materials. 
Therefore, the phonon modes under high pressure in Fig.~2(c) and Fig.~2(d) show the transition between the non-polar structures, which is from $Pm$\={3}$m$ to $R$\={3}$m$ at $P$ $\approx$ 30 GPa ($V$ $\approx$ 55 \AA$^3$). 
Our observations of the non-polar $R$\={3}$m$ structure at elevated hydrostatic pressures are similar to the cases observed in the (111)-oriented (and epitaxially strained) BTO films: there are experimental
evidences showing that the (111)-oriented BTO film has much smaller polarization than that of (001)-oriented (tetragonal) film.\cite{Nakagawara2002,Zhu2006} 
Oja \textit{et al}.~also reported that the (111) epitaxially strained BTO film can have a first order transition (having a jump in polarization) from $R$3$m$ to non-polar $R$\={3}$m$ by applying compressive-in-plane (anisotropic) stress.\cite{Oja2008} 
Similar to these reports, the observed non-FE state at elevated hydrostatic pressures can be ascribed to the strain origin. 
As pressure increases, the cell volume decreases monotonically, but the $c_{\rm H}/a_{\rm H}$-ratio (in hexagonal setting) varies non-linearly.
First, the $c_{\rm H}/a_{\rm H}$-ratio decreases continuously to $\sim$ 1.224  in $R$3$m$, and it remains unchanged within $Pm$\={3}$m$ symmetry below $P$ $\approx$ 30 GPa.
And then, it increases again as pressure increases above $P$ $\approx$ 30 GPa. 
In a restatement using the strain, the cell shrinks more along the $c_{\rm H}$-axis in $R$3$m$; it shrinks in an isotropic way in $Pm$\={3}$m$; it shrinks more along the $a_{\rm H}$-axis in $R$\={3}$m$. 
The cell is strained in an anisotropic manner in $R$\={3}$m$, which is a preferentially compressive-in-plane strain.
Therefore, we re-confirm that a strain effect prevents ferroelectricity by leading to a non-polar state in rhombohedral structure, similar to the Oja {\it et al}.\cite{Oja2008}
The difference is that the hydrostatic pressure in a BTO bulk induces successive second-order transitions (without a jump in polarization) in a sequence of $R$3$m$ (FE) - $Pm$\={3}$m$ (PE) - $R$\={3}$m$ (PE).

Thus far, we discussed the phonon modes at $\mathit{\Gamma}$-point.
To determine the full extent of structural stability, it is important to examine the phonon modes of all possible directions of the phonon-propagation vectors. 
The optical phonon modes were calculated for the propagation vectors pointing at the high-symmetry points of the Brillouin zone (BZ): the vectors were at $\mathit{\Gamma}$ (000), $X$ (010), $Z$ (001), $M$ (110), $R$ (011), and $A$ (111) for $P$4$mm$; and at $\mathit{\Gamma}$ (000), $X$ (001), $M$ (011), and $R$ (111) for both $R$3$m$ and $Pm$\={3}$m$. 
The above vectors are compatible with the 2$\times$2$\times$2 supercells used for the dispersion calculations at each of the cubic, tetragonal, and rhombohedral structures, i.e., the calculated phonon frequencies at these points are exact within numerical errors. 
Note that some characters for the BZ points in one structure deliver different meanings in another structure, as
will be shown below. For example, the $R$ point in $P$4$mm$, i.e. representing (011) point, corresponds to the $M$ point in $R$3$m$ or $Pm$\={3}$m$; the $A$ point for (111) in $P$4$mm$ to the $R$ point in $R$3$m$ or $Pm$\={3}$m$. 
To check the stability of the structures of $Pm$\={3}$m$, $P$4$mm$, and $R$3$m$ symmetries under ambient pressures, the optical phonon modes were calculated for the phonon-propagation vectors at the high-symmetry points of the BZ for selected pressures, $P$ = 0, 4, 14, 200, and 220 GPa. 
Fig.~4 summarizes the results calculated using the B1WC functional.

\begin{figure}[b]
\centering
\includegraphics[scale=0.5]{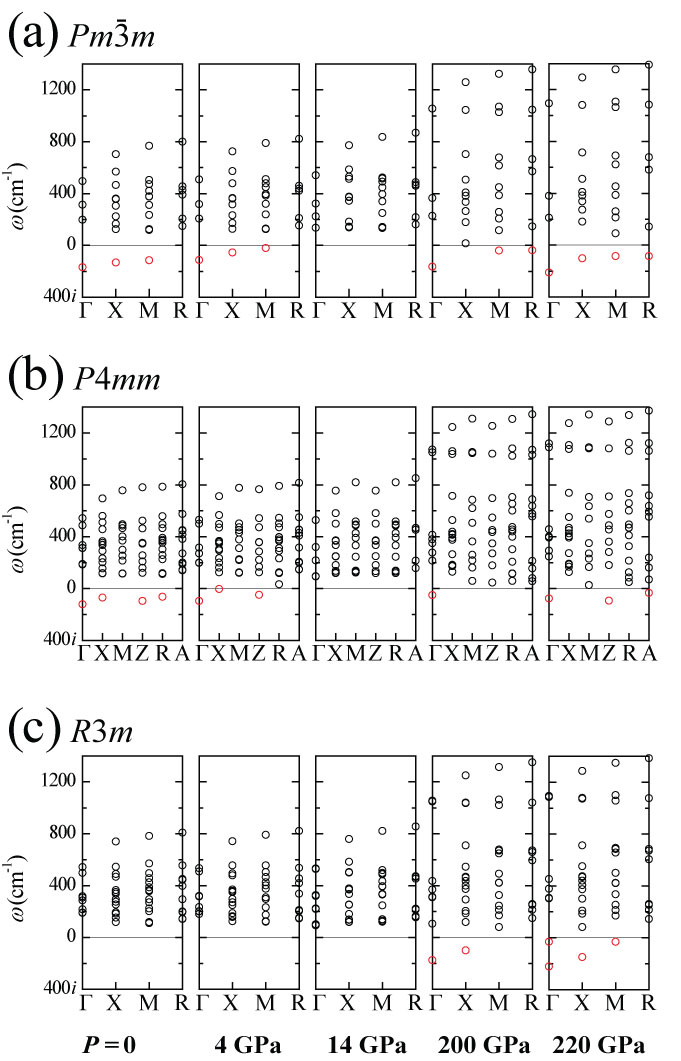}
\caption{Optical phonon modes calculated under hydrostatic pressures for the phonon-propagation vectors at the high-symmetry points of the BZ: of (a) $Pm$\={3}$m$ cubic for the vectors at $\mathit{\Gamma}$, $X$, $M$, and $R$; (b) $P$4$mm$ tetragonal for vectors at
$\mathit{\Gamma}$, $X$, $M$, $Z$, $R$, and $A$; of (c) $R$3$m$ rhombohedral for the vectors at $\mathit{\Gamma}$, $X$, $M$, and $R$. 
The modes with imaginary frequencies are plotted below zero. 
The modes were calculated for applied pressures of $P$ = 0, 4, 14, 200, and 220 GPa using the B1WC functional.}
\label{fig4}
\end{figure}

Let's first compare the results of three polymorphs under ambient pressure, i.e. $P$ = 0. 
At $P$ = 0, the $Pm$\={3}$m$ structure of BTO was found to be unstable from the calculated imaginary frequencies at $\mathit{\Gamma}$, $X$, and $M$, except at $R$, as shown in Fig.~4(a). 
This means that the cubic structure is unstable with respect to the phonon perturbations at three out of four high-symmetry points under consideration. 
The results for $P$4$mm$ (at $P$ = 0) also showed imaginary modes at $\mathit{\Gamma}$, $X$, $Z$, and $R$, but not at the $M$ or $A$ points. 
The imaginary phonon modes of $P$4$mm$, shown in Fig.~4(b), are clearly contrasted from the non-imaginary modes of $R$3$m$, as shown in Fig.~4(c). 
The $R$3$m$ (at $P$ = 0) has no imaginary frequencies for all high-symmetry points, which means that the structure is stable with respect to any dynamic perturbations. 
One might have predicted the stability of the $R$3$m$ from the findings of real-frequency-modes at the $R$ point in $Pm$\={3}$m$ and $A$ point in $P$4$mm$, both corresponding to the trigonal perturbations along the (111) directions.

Figure 4 summarizes the structural instabilities of the three polymorphs with respect to the dynamic perturbations as functions of pressure. Regarding the optical phonon modes of the $Pm$\={3}$m$ cubic at the high-symmetry points, shown in Fig.~4(a), the imaginary modes at the $\mathit{\Gamma}$, $X$, and $M$ points disappear gradually as pressure increases from zero to 14 GPa, and then the instabilities show up again at the $\mathit{\Gamma}$, $X$, $M$, and $R$ points under high pressures.
As is well-known, the triple-degenerate imaginary modes at the $\mathit{\Gamma}$-point represent the structural instability of $Pm$\={3}$m$, and they impart ferroelectric distortion into the $P$4$mm$ structure. 
Fig.~4(b) shows the cases of the $P$4$mm$ tetragonal. 
As pressure increases from zero, the number of imaginary frequency modes decreases to zero and then increases again at high pressures. 
The imaginary modes are removed from the $\mathit{\Gamma}$, $Z$, $X$, and $R$ points as the pressure increases from zero to 14 GPa, i.e. as the crystal symmetry relaxes from \textit{P}4\textit{mm} tetragonal to $Pm$\={3}$m$ cubic. 
The imaginary modes re-emerge at $\mathit{\Gamma}$, $Z$, and $A$ at higher pressures as the crystal
symmetry lowers itself from $Pm$\={3}$m$ cubic to $P$4$mm$ tetragonal. 
Fig.~4(c) shows the cases of the rhombohedral structure: the symmetries at each pressure are $R$3$m$ at $P$ = 0 and 4 GPa; $Pm$\={3}$m$ (cubic) at 14 GPa; $R$\={3}$m$ at 200 and 220 GPa.

The implications of the dynamic instabilities at $\mathit{\Gamma}$ (at zone center) are different from the other instabilities at high-symmetry points of the zone boundary. 
The $Pm$\={3}$m$ cubic BTO was well-known to have unstable phonon modes at the $\mathit{\Gamma}$, $X$, and $M$ points \cite{Ghosez1999}. 
One of the triple-degenerate unstable modes at $\mathit{\Gamma}$ induces a distortion along the
$z$-direction, which stabilizes the ferroelectric $P$4$mm$ BTO at least within the five-atom unit cell description. 
The remaining two unstable modes at $\mathit{\Gamma}$ in $P$4$mm$ are the instabilities along the $x$- and $y$-directions, being suggestive of additional antiferroelectric distortions involving the unit-cell-doubling. 
The instabilities at the $X$ point of $Pm$\={3}$m$ are doubly-degenerate, and their eigenvectors are dominated by the bending motions of titanium and apical oxygen at $P$ = 0 and by the buckling motions of titanium and planar oxygen's at $P$ = 220 GPa. 
According to Bousquet \textit{et al}.'s study on (BaTiO$_3$)$_m$/(BaO)$_n$ superlattices,\cite{Bousquet2010} the centosymmetric tetragonal $P$4/$mmm$ structure becomes dynamically unstable as $m/n$ exceeds 4/2: $X$ point instability appears first, which is related to the antiferroelectric distortion, and as $m/n$ is increased, the instability extends to $M$ point, but both
instabilities are non-polar modes. 
In addition, the $M$/$R$ point instabilities in $Pm$\={3}$m$ structure are related to structural distortions accompanying the in-phase/out-of-phase rotations of the alternating oxygen octahedra, which were observed in the SrTiO$_3$\cite{Xie2008} and EuTiO$_3$\cite{Rushchanskii2012} dynamically-stabilized into the $I$4/$mcm$ structure. 
There are similar but controversial reports in BaZrO$_3$,\cite{Bilic2009} in which the dynamic instabilities at the $M$ and $R$ point are stabilized into the larger non-polar structure with the lower symmetry.

\section{Conclusion}
We studied the phonon modes of three polytypes of BaTiO$_3$, $Pm$\={3}$m$, $P$4$mm$, and $R$3$m$, by \textit{ab}-\textit{initio} calculations using DFT, HF, and hybrid functionals. 
The zone-center optical phonon modes calculated under zero-external pressure, i.e. with the fully relaxed geometries, were compared systematically using eight different functionals.
The B1WC hybrid functional predicted the closest values to the LDA results, except for the displaced $A_1$ TO mode at $\sim$ 300 cm$^{-1}$, but the pressure-dependence of the phonon mode was surprisingly similar to each other. 
The evolution of the phonon-branches as functions of pressure revealed two successive transitions, each
preceded by the softening of modes, the FE ($P$4$mm$) to PE ($Pm$\={3}$m$)
transition at below $\sim$ 10 GPa and the re-entrant PE to FE transition at above $\sim$ 150 GPa. 
In the rhombohedral structure, the phonon-branching behavior and missing polarization at pressures above 30 GPa suggested the re-entrant PE ($Pm$\={3}$m$) to the PE ($R$3$m$) transition. 
An analysis of the phonon modes propagating to the high-symmetry directions under high pressure, where the
ferroelectricity recurs, suggested that the $Pm$\={3}$m$ phase has ferroelectric-distortive
instability at $\mathit{\Gamma}$ and non-polar instability at $X$, $M$, and $R$.

\section*{Acknowledgments} 
This study was supported by the National Research Foundation of Korea (NRF) grant funded by the Ministry of
Education, Science and Technology (MEST), No.~2012006641. 
The computation is supported by the Korea Institute of Science and Technology Information (KISTI) Supercomputing Center through contract No.~KSC-2012-C2-36.

\newpage
\clearpage

\begin{widetext}
{\Large\bf\underline{Supplement materials}}
{
\begin{itemize}
   \item Polarizations and structural parameters as functions of pressure
     \begin{itemize}
       \item Supplement Fig. 1: For P4mm calculated by using LDA
       \item Supplement Fig. 2: For R3m calculated by using LDA
       \item Supplement Fig. 3: For P4mm calculated by using B1WC
       \item Supplement Fig. 4: For R3m calculated by using B1WC
     \end{itemize}
   \item Optical phonon modes calculated under pressures for phonon-propagation vectors at high-symmetry points of the BZ
     \begin{itemize}
       \item Supplement Fig. 5: For Pm-3m, P4mm, and R3m calculated by using LDA
     \end{itemize}
\end{itemize}
} 
\end{widetext}

\vbox{}
\vspace*{0cm}

\vbox{
\begin{center}
\includegraphics[scale=0.5]{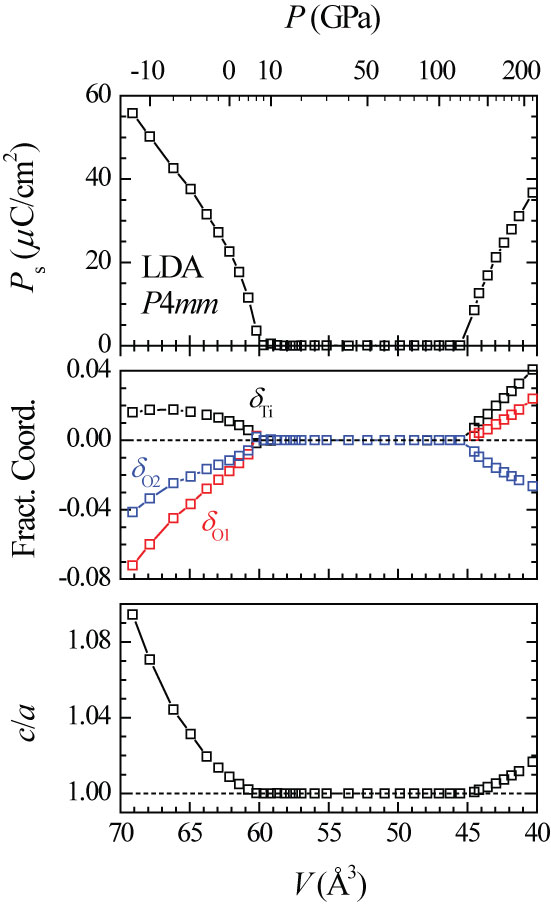}
\end{center}
{\small {\bf Supplement Figure 1:} \\
Polarizations and structural parameters of $P$4$mm$ as functions of pressure: (a) polarizations, (b) fractional coordinates, and (c) $c/a$-ratios. LDA functional was used. Note that the fractional coordinates of the basis atoms in $P$4$mm$ are as follows: Ba (0, 0, 0), Ti (1/2, 1/2, 1/2 + $\delta$Ti), O1 (1/2, 1/2, $\delta$O$_1$), and O$_2$ (1/2, 0, 1/2 + $\delta$O$_2$).}
}

\vbox{}
\vspace*{0cm}

\vbox{
\begin{center}
\includegraphics[scale=0.5]{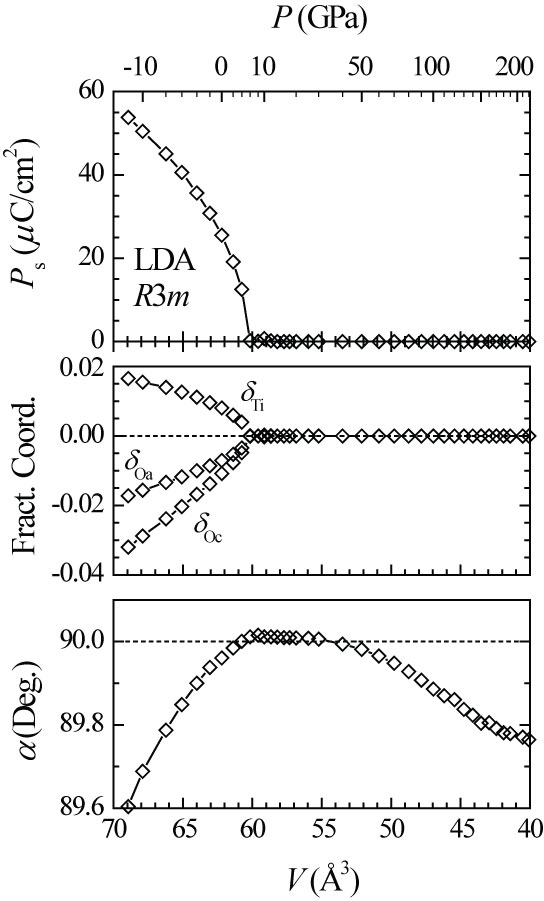}
\end{center}
{\small {\bf Supplement Figure 2:} \\
Polarizations and structural parameters of $R$3$m$ as functions of pressure: (a) polarizations, (b) fractional coordinates, and (c)  rhombohedral angle $\alpha$'s. LDA functional was used. Note that the fractional coordinates of the basis atoms in $R$3$m$ are as follows: Ba (0, 0, 0), Ti (1/2 + $\delta$Ti, 1/2 + $\delta$Ti, 1/2 + $\delta$Ti), and O (1/2 + $\delta$O$_c$, 1/2 + $\delta$O$_c$, $\delta$O$_a$ + $\delta$O$_c$).}
}

\vbox{}
\vspace*{4cm}

\vbox{
\begin{center}
\includegraphics[scale=0.5]{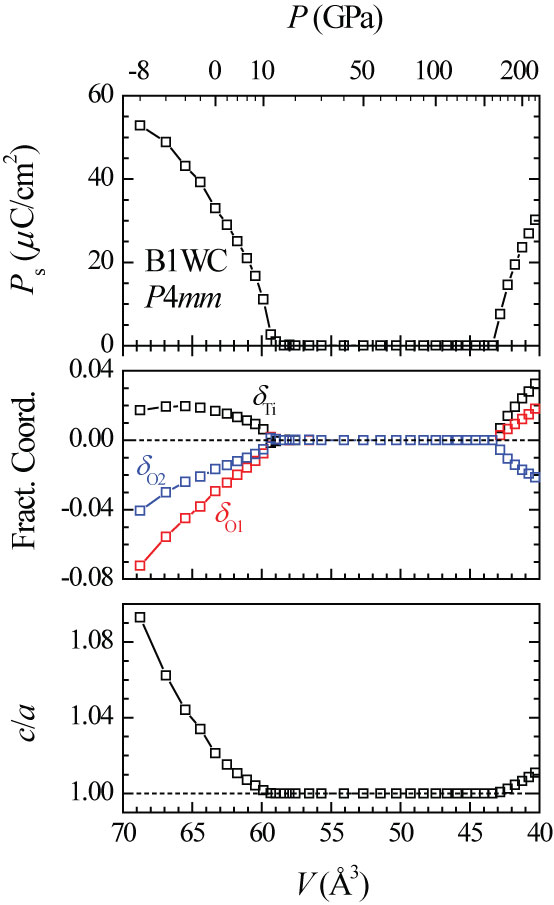}
\end{center}
{\small {\bf Supplement Figure 3:} \\
Polarizations and structural parameters of $P$4$mm$ as functions of pressure: (a) polarizations, (b) fractional coordinates, and (c) $c/a$-ratios. B1WC functional was used. Note that the fractional coordinates of the basis atoms in $P$4$mm$ are as follows: Ba (0, 0, 0), Ti (1/2, 1/2, 1/2 + $\delta$Ti), O1 (1/2, 1/2, $\delta$O$_1$), and O$_2$ (1/2, 0, 1/2 + $\delta$O$_2$).}
}

\vbox{}
\vspace*{4cm}

\vbox{
\begin{center}
\includegraphics[scale=0.5]{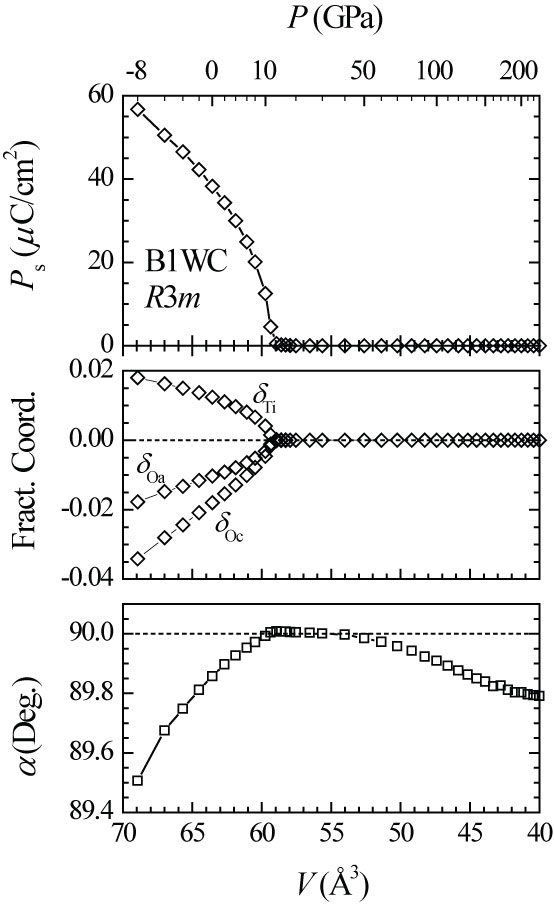}
\end{center}
{\small {\bf Supplement Figure 4:} \\
Polarizations and structural parameters of $R$3$m$ as functions of pressure: (a) polarizations, (b) fractional coordinates, and (c) rhombohedral angle $\alpha$'s. B1WC functional was used. Note that the fractional coordinates of the basis atoms in $R$3$m$ are as follows: Ba (0, 0, 0), Ti (1/2 + $\delta$Ti, 1/2 + $\delta$Ti, 1/2 + $\delta$Ti), and O (1/2 + $\delta$O$_c$, 1/2 + $\delta$O$_c$, $\delta$O$_a$ + $\delta$O$_c$).}
}

\newpage
\clearpage


{\widetext
\vbox{}
\vspace*{1cm}
\begin{center}
\includegraphics[scale=0.7]{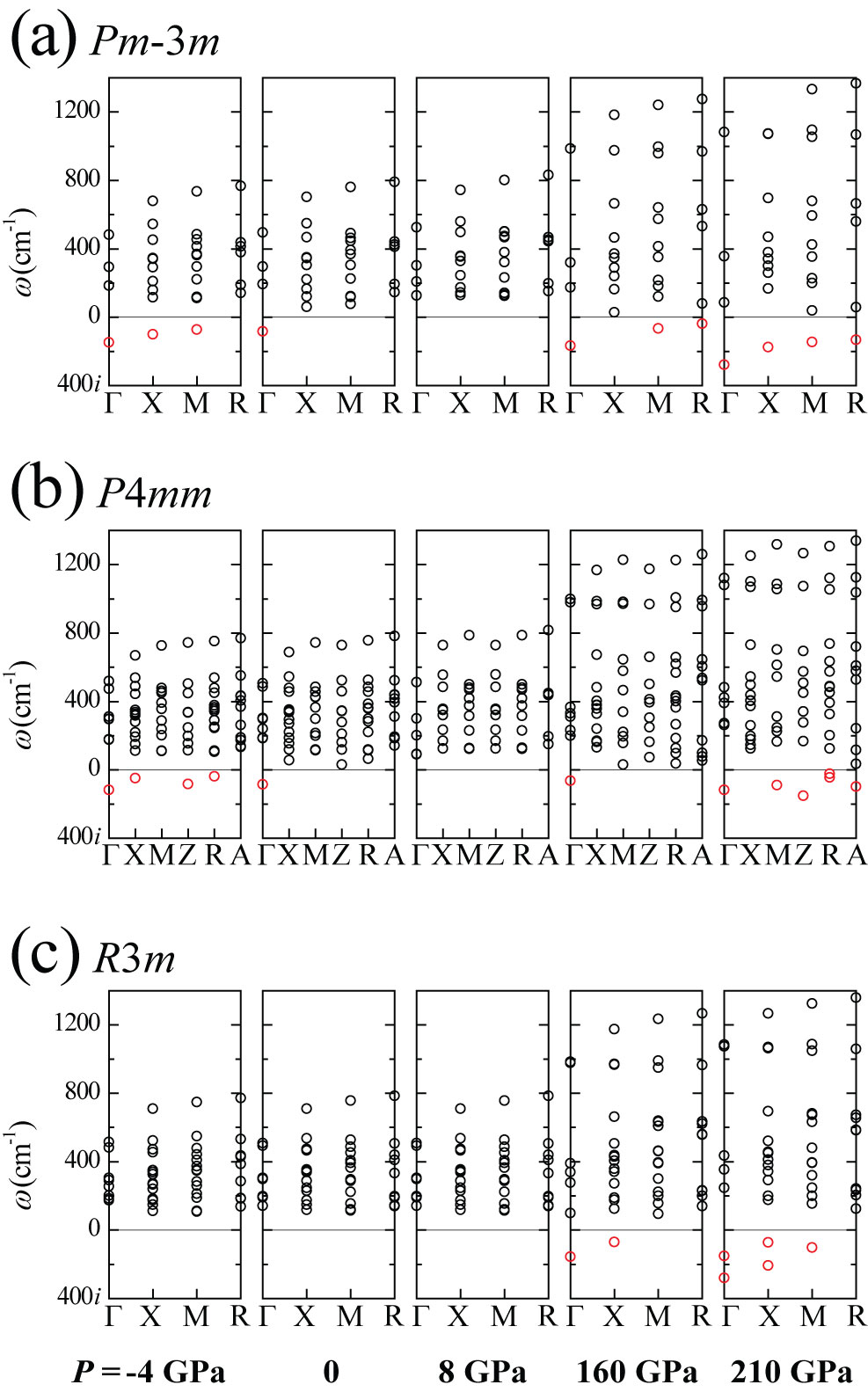}
\end{center}
{\small {\bf Supplement Figure 5:} \\
Optical phonon modes calculated under hydrostatic pressures for the phonon-propagation vectors at the high-symmetry points of the BZ: of (a) $Pm$\={3}$m$ cubic for the vectors at $\mathit{\Gamma}$, $X$, $M$, and $R$; (b) $P$4$mm$ tetragonal for vectors at $\mathit{\Gamma}$, $X$, $M$, $Z$, $R$, and $A$; of (c) $R$3$m$ rhombohedral for the vectors at $\mathit{\Gamma}$, $X$, $M$, and $R$. 
The modes with imaginary frequencies are plotted below zero. 
The modes were calculated for applied pressures of $P$ = -4, 0, 8, 160, and 210 GPa by using the LDA functional.}
}

\end{document}